\documentclass[aps,prd,twocolumn,showpacs,amsmath,amssymb]{revtex4}
\usepackage{graphicx}
\usepackage{dcolumn}
\usepackage{bm}
\usepackage{amsmath}    
\usepackage{amsfonts}
\usepackage{amssymb}
\def\beqra{\begin{eqnarray}}
\def\eeqra{\end{eqnarray}}
\def\beq{\begin{equation}}
\def\eeq{\end{equation}}

\def\vp{\varphi}

\def\D{\nabla^2}

\def\p{\partial}

\def\de{\delta}

\def\a{\alpha}
\def\b{\beta}
\def\n{\nu}
\def\m{\mu}

\def\mh{\mathcal{H}}
\def\mr{\mathcal{R}}

\def\mp{\mathcal{P}}

\def\fr{\frac}
\def\pr{\prime}
\def\r{\rho}
\def\tx{\tilde{x}}

\begin{document}
\title{Stochastic gravitational wave background from cold dark matter
halos}
\author{Carmelita Carbone}
\email{carbone@sissa.it}
\affiliation{SISSA/ISAS, Astrophysics Sector, Via Beirut 4, I-34014, Trieste,
  Italy and \\ 
  INFN, Sezione di Trieste, Via Valerio, 2, 34127, Trieste, Italy}
\author{Carlo Baccigalupi}
\email{bacci@ita.uni-heidelberg.de}
\affiliation{Institut f$\ddot{\rm u}$r Theoretische Astrophysik,
  Universit$\ddot{\rm a}$t Heidelberg, Tiergartenstrasse 15, D-69121,
  Heidelberg, Germany, \\
  SISSA/ISAS, Astrophysics Sector, Via Beirut 4, I-34014, Trieste, 
  Italy and \\
  INFN, Sezione di Trieste, Via Valerio, 2, 34127, Trieste, Italy}
\author{Sabino Matarrese}
\email{matarrese@pd.infn.it}
\affiliation{Dipartimento di Fisica `Galileo Galilei', Universit\`a di 
  Padova and \\ 
  INFN, Sezione di Padova, Via Marzolo 8, I-35131 Padova, Italy}

\date{\today}

\begin{abstract}
The current knowledge of cosmological structure formation suggests that 
Cold Dark Matter (CDM) halos possess a non-spherical density profile, implying 
that cosmic structures can be potential sources of gravitational waves 
via power transfer from scalar perturbations
to tensor metric modes in the non-linear regime.
By means of a previously developed mathematical formalism
and a triaxial collapse model, we numerically estimate the
stochastic gravitational-wave background generated by CDM halos during the 
fully non-linear stage of their evolution.  
Our results suggest that the energy density associated with this background
is comparable to that produced by primordial tensor modes at frequencies 
$\n\approx10^{-18}-10^{-17}$Hz if the energy scale of 
inflation is $V^{1/4}\approx 1-2\times 10^{15}$ GeV, and that these
gravitational waves could give rise to several cosmological effects, 
including secondary CMB anisotropy and polarization.
\end{abstract}

\maketitle

\section{Introduction}

Sources of gravitational waves (GW) are commonly separated in two types: 
astrophysical and cosmological. 

The first kind of sources can produce a stochastic background which 
provides interesting information on the distribution of compact objects 
at relatively low redshifts, such as star formation and supernova rates, 
black-hole growth mechanisms and other important phenomena. 
Such a background is generated by neutron stars, black holes
and the associated binary systems, which emit in the frequency range
$\n_e\approx 10^2-10^4$Hz. (e.g \cite{Maggiore,Ferrari et al. 1999}), or by
galactic merging of unresolved binary white dwarfs with frequencies in the
range $\n_e\approx 10^{-4}-10^{-2}$Hz \cite{Postnov97, Kosenko98, Schneider et
al. 2000, Schneider et al. 2001, Farmer and Phinney 2003}. 

Besides binary systems of super-massive black holes in the galaxy center,
which could emit at $\n_e\approx10^{-4}$Hz, hence detectable by LISA 
({\it e.g.} Ref.~\cite{Grishchuk03}), 
the principal example of gravitational waves of
cosmological origin is represented by the relic radiation which has been 
generated by quantum fluctuations of the metric tensor 
during the inflationary era. 
The detection of this relic background would shed light on 
the physics of the very early Universe, since its strain 
amplitude is proportional to the square of the inflation 
energy scale. 
Primordial backgrounds can be generated by various mechanisms and are 
characterized by a large frequency interval which extends from 
a few $10^{-18}$ Hz to a few GHz, allowing their detection by 
markedly different ways of observation \cite{Maggiore}.

One of the best strategies for detecting the relic gravitational 
radiation is to exploit the imprints it leaves on the Cosmic Microwave 
Background (CMB) temperature anisotropy and polarization 
\cite{Hu et al. 1998, reviews, books}. 
More specifically, the CMB photons are very sensitive to
primordial GWs with frequencies $\approx10^{-17}$Hz, 
which correspond to the comoving size of the Hubble radius at last scattering,
when tensor metric modes, being damped by the horizon entering, produce the
largest amount of temperature quadrupole anisotropy and, consequently, by
Thomson scattering, the largest amount of polarization
\cite{Pritchard&Kamionkowski04}. 
It may be shown that the curl component in the polarization pattern, commonly 
known as B-mode, is excited by vector and tensor cosmological perturbations
only; therefore, if initial fluctuations are created very early, {\it e.g.}
during inflation so that the vector growth is damped, \emph{primary} B-modes
can be produced only by tensor perturbations and, therefore, a possible
detection will represent the incontrovertible proof of their existence 
\cite{Kamionkowski et al. 1997, Zaldarriaga&Seljak97, Seljak&Zaldarriaga97}.

Unfortunately, there are mechanisms that can produce
\emph{secondary} B-modes, the principal one being represented by the
gravitational lensing, {\it i.e.} \emph{cosmic shear} (CS)
\cite{Bartelmann&Schneider01}, 
which distorts the primary CMB pattern, in particular converting E- into
B-modes \cite{Zaldarriaga&Seljak98}. 
Luckily, although comparable, 
B-modes from primordial GW exhibit their peak at multipoles $l\approx 100$, 
corresponding to the degree scale, while, for lensed B-modes, the peak is at 
$l\approx1000$, corresponding to the arcminute scale. 
Nonetheless, if the energy scale of inflation is $V^{1/4}\le
2-4\times10^{15}$GeV, the CS-induced curl is a foreground for the
$l\approx50-100$ primordial GW-induced B-polarization.
This important contamination has to be removed
in order to detect relic gravitational waves \cite{seljak_hirata_2004}.
However, as suggested by WMAP measurements, early reionization should produce
a large-angle bump in the primordial GW-induced B-modes, allowing 
a possibly easier detection, without confusion with the CS-induced curl 
\cite{Cabella&Kamionkowski04}.

Actually, besides gravitational lensing, 
for cosmological models which constantly seed fluctuations in the
geometry, {\it e.g.} topological defects, vector metric perturbations 
can be huge and can produce non-negligible effects on the CMB photons as, 
in particular, B-mode polarization, unlike to what
happens in inflationary models \cite{Turok et al. 1998}. 
On the other hand, no relevant contribution from these objects is 
indicated by the modern cosmological probes. 

In the present paper, we are interested in the cosmological stochastic GW
background produced by Cold Dark Matter (CDM) halos via power transfer
from scalar and possible vector perturbations to tensor metric modes, during
the strongly non-linear stage of their evolution \cite{CM}. It differs from
other cosmological backgrounds, as that produced during the mildly non-linear
stage \cite{mm2}, since density and velocity fields can be, in this case, 
highly non-linear. 

Since the non-linear evolution of CDM halos occurs on a cosmological 
timescale, the produced gravitational radiation may be relevant at 
frequencies comparable to those of the primordial GW which affect the 
CMB photons and, therefore, can produce secondary CMB anisotropy and 
polarization, expecially B-modes, that could represent a foreground for 
the detection of the relic radiation. 
 
Moreover, as for the case of black holes and neutron stars, 
the analysis of the stochastic background produced by 
highly non-linear cosmic structures, could bring information on 
their distribution, evolution, shape and composition, shedding light on 
many open issues.

The plan of paper is as follows. In Sec.~II we briefly outline the
mathematical formalism and show the analytical formulas we use to estimate the
GW output from CDM halos. In Sec.~III we introduce the homogeneous ellipsoid 
dynamics as an approximation to the halo virialization and adopt the halo
mass function of Ref.~\cite{Sheth et al. 2001} to describe their
distribution. In Sec.~IV we explain the technique
used for the numerical evaluation of the stochastic GW background, while 
in Sec.~V we show and discuss our results. Finally Sec.~VI contains our 
concluding remarks. 

\section{Gravitational Radiation: \newline basic equations}

The evolution of cosmological perturbations away from the
linear regime is rich of several effects, such as the mode-mixing 
of different types of fluctuation, which 
not only implies that different Fourier modes influence each other, but also
that density perturbations act as a source for curl vector modes and
gravitational waves.

Accordingly, cosmic structures can generate tensor metric
modes during the non-linear stage of their evolution and, in
particular, this mechanism applies 
to dark matter halos around galaxies and galaxy clusters in the highly
non-linear regime.

In the present paper, adopting the mathematical formalism developed 
in Ref.~\cite{CM}, we estimate the output in gravitational waves from
cosmic structures, following their evolution from the linear to the
highly non-linear level. 
More specifically, the evaluation of this gravitational radiation 
is possible on scales much larger than the Schwarzschild radius of collapsing
bodies, by means of a ``hybrid approximation'' \cite{CM} 
of the Einstein field equations, which mixes post-Newtonian (PN) 
({\it e.g.} \cite{ch, ch2, ch3, tomita3, tomita4, sa, mt}) 
and second-order perturbative techniques ({\it e.g.} 
\cite{pyne, mm, mmb, tomita, mps, tomita2})  
to deal with the perturbations of matter and geometry. 
This approach gives a more accurate description of gravitational waves 
generated by non-linear
CDM structures than the standard second-order perturbation theory \cite{mm2}, 
which can only account for small deviations from the linear regime, 
or the Newtonian quadrupole radiation \cite{grav,landau}; indeed, it upgrades
the weak-field limit of Einstein equations to account for PN scalar
and vector metric perturbations and for leading-order source terms of
metric tensor modes. 
It provides, on small scales, a PN approximation to the source of
gravitational radiation, and, on large scales, it converges to the first
and second-order perturbative equations as obtained {\it e.g.} 
in Ref.~\cite{mhm}, but still describing, on all the cosmologically 
relevant scales, the dynamics of the involved CDM structures 
by means of the standard Newtonian Poisson, Euler and continuity
equations ({\it e.g.} \cite{peebles})
\begin{align} 
\label{Poisson}
\nabla^{2}\varphi = 4\pi Ga^{2} \delta\rho\;,
\end{align}
\begin{align}
\label{continuity}
\rho^{\pr}+3\mathcal{H}\rho+\partial_{\n}(\rho\,v^{\n})=0\;,
\end{align}
\begin{align}
\label{Euler}
v_{\a}^{\pr}+\mathcal{H} v_{\a}+v_{\n}\partial^{\n}v_{\a
}=-\partial_{\a}\varphi\;,
\end{align}
where $\vp$ is the gravitational potential associated with the density 
perturbation, $\r=\bar{\r}+\de \r$ is the total matter density composed by
the background matter density, $\bar{\r}$, 
and the matter density perturbation,
$\de \r$, and, finally, $\mathbf{v}$ is the peculiar velocity field associated
to the CDM halos. 
Greek indices denote spatial components; we adopt
conformal time $\eta$ and comoving coordinates $x^\a$, 
in the Poisson gauge, and assume that the Universe is 
spatially flat and filled with a cosmological constant 
$\Lambda$ and a pressureless fluid whose
stress-energy tensor reads $T^i{}_j=\r u^i u_j$ ($u^i u_j=-1$).
Finally, $\mh \equiv a^\pr/a$, where primes indicate differentiation with
respect to $\eta$ and $a$ is the scale factor of the Universe which evolves
according the Friedmann-Robertson-Walker background model. 

Indeed, as the background cosmology, we have adopted a flat $\Lambda$CDM 
model with present baryon density given by $\Omega_{0\rm b}=4.318 \times 
10^{-2}$, dark and CDM energy density $\Omega_{0\Lambda}=0.7434$, 
$\Omega_{0\rm {CDM}}=0.2134$, Hubble constant $H_{0}=100h$ km/sec/Mpc 
where $h=0.7199$ and three massless neutrino species; the 
primordial perturbation spectrum is made by scalars only, normalized 
by $\sigma_8=\sigma(R=8\,h^{-1}\,\rm{Mpc})=0.9$, with spectral index 
$n_s=0.96$ \cite{Tegmark et al. 2004, Spergel et al. 2003}.

Accordingly to Ref.~\cite{CM}, in order to evaluate 
the stochastic background of gravitational radiation generated by CDM halos,
we will exploit the formula  expressing the solution of the inhomogeneous GW
equation on scales well inside the Hubble horizon and in the so-called wave
zone, which is 
\begin{align} 
\label{hab_quadrupole}
h^\a{}_\b(\eta,\mathbf{x})=
\fr{4G}{c^4}\fr{1}{ar}\mp^{\a\;\;\m}_{\;\;\n\;\;\;\b}
\left[a^3\int d^3\tx\, \mr^\n_{\textrm{eff}\,\m}\right]_{ret} \;, 
\end{align} 
where $r$ is the comoving distance between source 
and observer while the projection operator is given 
by $\mp^\a{}_\b\equiv\de^\a{}_\b-x^\a x_\b/r^2$. 
Eq.~(\ref{hab_quadrupole}) expresses the GW output 
$h^\a{}_\b$ in terms of integrals over 
the source ``stress distribution'' $\mr^\a_{\textrm{eff}\,\b}$, given by 
\begin{align}
\label{Rab_eff}
\mr^\a_{\textrm{eff}\,\b}
&=\r\left(v^\a v_\b-\fr{1}{3} v^2 \,\de^\a{}_\b \right)+\nonumber\\
&+\fr{1}{4\pi Ga^2}\left(\p^\a\vp\, \p_\b\vp
-\fr{1}{3}\p^\n\vp\,\p_\n\vp\,\de^\a{}_\b \right)\;. 
\end{align}
The subscript ``{\it ret}'' in Eq.~(\ref{hab_quadrupole})
means that the quantity has to be evaluated at the retarded 
space-time point $\left(\eta-r/c,\mathbf{\tx}\right)$, 
{\it i.e.} at the source and at the emission time. 

\section{The ellipsoidal collapse model}

Recently, N-body simulations in CDM models have shown departure 
of the halo density profile from the spherical symmetry 
({\it e.g.} \cite{Jing&Suto2000}) and suggest a triaxial shape which 
seems to be confirmed by optical, X-ray and lensing observations of 
galaxy clusters ({\it e.g.} \cite{West, Plionis et al.1991}).
 
Consequently, according to the arguments in the previous Section, 
CDM halos are potential sources of gravitational radiation 
through power injection from the gravitational potential and 
peculiar velocity, especially during the highly non-linear stage of their
evolution, when density contrasts and velocity fields can be strongly
non-linear. 
Since the aim of this paper is to evaluate the stochastic GW
background generated by a distribution of cosmic structures, 
in this Section we will describe the model adopted to approximate their
dynamics and virialization.

\subsection{The homogeneous ellipsoid dynamics}
We will use the gravitational collapse of homogeneous ellipsoids as described
in Ref.~\cite{B&M},
which developed a picture of cosmic structure formation that
identifies virialized cosmological objects with peak patches in the initial
Lagrangian space. These peaks represent overdensities in the initial Gaussian
density field whose evolution is approximated by a 
homogeneous ellipsoid dynamics. Each
perturbation evolves under the influence of its own gravity and under the
external tidal field (generated by the surrounding matter) which, together with
initial conditions, is chosen to reproduce the Zel'dovich approximation in the
linear regime. Virialization is defined as the time when the third axis
collapses and, following Ref.~\cite{B&M}, each axis is frozen once it 
has reached a freeze-out radius, chosen so that the density contrast at
virialization, in the limit of spherical collapse, is the same as 
prescribed by the top-hat model. 

The peculiar velocity field is conveniently described in the system 
identified by the three principal axes, characterized by three 
different scale factors $R_\alpha$ ($\alpha =1,2,3$); thus, inside the
homogeneous ellipsoid, peculiar velocities may be written as
\begin{align}
\label{velocity}
v_\a=\left(a\fr{\dot{R_\a}}{R_\a}-\dot{a}\right)x_\a \;,
\end{align}
where we are still adopting comoving coordinates but now time derivatives 
are with respect to the proper time $dt=a\, d\eta$. 

The internal peculiar gravitational potential, still with respect to the
principal-axis system, is given by (see Ref.~\cite{B&M} for details)
\begin{align}
\label{potential}
\vp=
\pi Ga^2\bar{\r}\left[\sum_{\a=1}^3 (\de b_\a+2\lambda^\pr_\a)x_\a^2\right] \;,
\end{align}
where $\de \equiv \de\r/\bar{\r}$ is the matter density contrast
while the factors $b_\a$ are given by
(see {\it e.g.} Refs.~\cite{Chandra69, binney})
\begin{align}
\label{factors}
b_\a=R_1 R_2 R_3 \int_0^\infty \fr{ds}{\left(R_\a^2+s\right)
\sqrt{\left(R_1^2+s\right)\left(R_2^2+s\right)\left(R_3^2+s\right)}}
\end{align}
Finally, the coefficients $4\pi Ga^2\bar{\r}\lambda^\pr_\a$ are the
eigenvalues of the traceless external tidal tensor (proportional to the
traceless part of the peak strain) for which a linear approximation is
assumed \cite{B&M}, imposing that it evolves through the same equations 
satisfied by the linear growth-factor of density fluctuations in the 
considered cosmological background.  

After imposing the Zel'dovich approximation to fix the initial 
conditions on the proper ellipsoid axis lengths and their time derivatives, 
the evolution of an ellipsoidal perturbation is specified through the
equations \cite{B&M} 
\begin{align}
\label{axis}
\fr{d^2R_\a}{dt^2}=\fr{\Lambda c^2}{3}R_\a-4\pi G\bar{\r} R_\a 
\left(\fr{1}{3}+\fr{\de}{3}+\fr{b_\a^\pr}{2}\de+\lambda^\pr_\a \right) \;,
\end{align}
\begin{align}
\label{background}
\fr{d^2a}{dt^2}=\left(-\fr{4\pi G}{3}\bar{\r}+\fr{\Lambda c^2}{3}\right)a \;,
\end{align}
\begin{align}
\label{const1}
\r R_1 R_2 R_3=\textrm{const} \;,
\end{align}
\begin{align}
\label{const2}
\bar{\r}a^3=\textrm{const} \;,
\end{align}
where, in Eq.~(\ref{axis}), $b_\a^\pr=b_\a-2/3$.

Eqs.~(\ref{axis})-(\ref{const2}) have been numerically integrated using 
a fourth-order Runge-Kutta scheme and the integrals (\ref{factors}) 
have been evaluated by means of the so-called Carlson's elliptic function of
the third kind.  

\begin{figure}[tbp]
\includegraphics[width=0.48\textwidth]{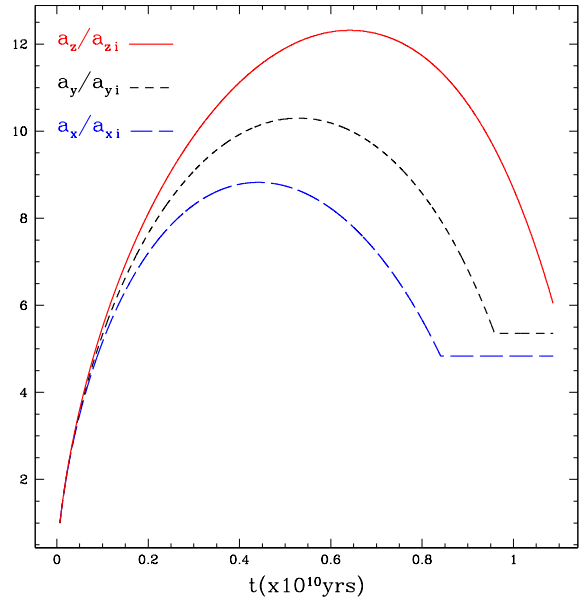}
\caption{Evolution of the principal axis scale factors for the most probable
  ellipsoid of mass $M=5\times 10^{15} M_\odot$ and initial overdensity
  $\de(z_i=40)=6.4\times 10^{-2}$, embedded in a flat $\Lambda$CDM
  universe, at distance $D=100\;{\rm Mpc}$ from the observer.} 
\label{collapse}
\end{figure}

Fig.~\ref{collapse} shows the axis evolution versus time 
of a homogeneous ellipsoid of mass $M=5\times 10^{15} M_\odot$ and initial
overdensity $\de(z_i=40)=6.4\times 10^{-2}$ at comoving 
distance $D=100\;{\rm Mpc}$ from the observer. 
The shape of the ellipsoid is the most probable in terms 
of the distribution of ellipticity and prolateness, 
to be defined in the next sub-Section. 
The evolution follows Eqs.~(\ref{axis}-\ref{const2}) 
and the axis freezing out method suggested in Ref.~\cite{B&M}. 
Let us stress that, contrary to other ellipsoidal collapse
schemes ({\it e.g.} \cite{W&S}), this model implies that
virialization is reached when the third and not the first axis collapses, 
while the freezing out method avoids $\de\longrightarrow\infty$. 

In order to estimate the GW output by CDM cosmic structures,
we insert Eqs.~(\ref{velocity})-(\ref{potential}) in
Eqs.~(\ref{hab_quadrupole})-(\ref{Rab_eff}) during the collapse of
each homogeneous ellipsoid which represents, in our simulation, a CDM halo
evolving towards virialization. 

\begin{figure}[tbp]
\includegraphics[width=0.48\textwidth]{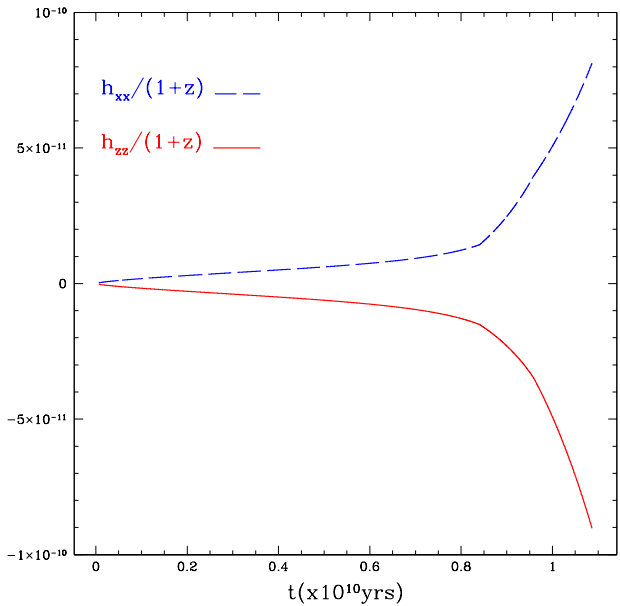}
\caption{Two of the three non-vanishing traceless source 
components generated by the halo collapse of Fig.~\ref{collapse}}
\label{gws}
\end{figure}

In Fig.~\ref{gws} we show two of the three non-vanishing traceless source 
components generated by the halo collapse of Fig.~\ref{collapse}. 
These components are evaluated with respect to the eigenframe 
of the ellipsoid principal axes at rest with respect to the expanding
cosmological background; by performing a transverse projection,
the gravitational waves in the observer frame are obtained.
Actually, Fig.~\ref{gws} represents these two components
divided by $1+z=1/a(t)$, in order to separate the effects of the background 
expansion, included in Eq.~(\ref{hab_quadrupole}), 
from the halo evolution itself.


\subsection{The most probable ellipsoid and the halo mass function}

Once the cosmological background model is fixed, 
the evolution of an ellipsoidal
perturbation is determined by three parameters given by the three initial 
eigenvalues of what, in the Zel'dovich approximation, 
is called the deformation tensor, $d_{\a\b}=(1/a^2)\D_\a\D_\b\vp$; 
the latter are related to the initial ellipticity \emph{e}, prolateness
\emph{p} and linear density contrast $\de$ of the perturbation; 
those relations read \cite{Sheth et al. 2001}
\begin{align}
\label{ellipticity}
e=\fr{\lambda_1-\lambda_3}{2\de} \;,
\end{align}
\begin{align}
\label{prolateness}
p=\fr{\lambda_1+\lambda_3-2\lambda_2}{2\de} \;,
\end{align}
\begin{align}
\label{density}
\de=\lambda_1+\lambda_2+\lambda_3 \;,
\end{align}
where the $\lambda_\a$ are the eigenvalues of $d_{\a\b}$ with 
$\lambda_1\ge\lambda_2\ge\lambda_3$, which, if $\de\ge0$, implies $e\ge0$  
and $-e\le p\le e$. 

For a Gaussian random density field, smoothed in real space with a 
top-hat filter of size $V=4\pi R^3/3$ and mass $M=4\pi\bar{\r} R^3/3$, 
on average and for a given $\de$ the prolateness is $p=0$; consequently, 
the most probable ellipticity is $e_{mp}=(\sigma /\de)/\sqrt{5}$. Here 
$\sigma=\sigma(R)$ represents the linear rms value of the $\de$ distribution 
\cite{Sheth et al. 2001}.
 
From these considerations and from the homogeneous ellipsoid collapse model as
described in Ref.~\cite{B&M}, the authors of Ref.~\cite{Sheth et al. 2001} have
determined the shape of the moving barrier, {\it i.e.} the critical overdensity
required for CDM structure virialization at redshift $z$; that is 
\begin{align} 
\label{barrier}
B(\sigma^2,z)=\sqrt{q}\de_{sc}(z)\left[1+\b(\a\nu)^{-\a}\right] \;,
\end{align}
where $\nu\equiv\left[\de_{sc}(z)/\sigma(M)\right]^2$, $\de_{sc}(z)$ is the
critical overdensity required for spherical collapse at $z$ extrapolated using
linear theory to the present time, and $\sigma$ is the linear rms value of the
initial density fluctuation field also extrapolated to the present
time. The parameters $\b\approx0.485$ and $\a\approx0.615$ come from
ellipsoidal dynamics and the value  $q\approx0.75$ comes from normalizing
the model to simulations \cite{Sheth&Tormen 2002}.

Using Eq.~(\ref{barrier}) in the excursion set approach in order to obtain the
distribution of the first crossings of the barrier by independent random
walks, the authors of Refs.~\cite{Sheth et al. 2001,Sheth&Tormen99} have
derived the 
average comoving number density of halos of mass $M$, {\it i.e.} the so-called
unconditional halo mass function  
\begin{align}
\label{MF}
n(M,z)dM =& \sqrt{\frac{2qA^2}{\pi}}\,
\frac{\rho_0}{M^2}\,\frac{\delta_{sc}(z)}{\sigma(M)}\,
\left[1+\left(\frac{\sigma(M)}{\sqrt{q}\delta_{sc}(z)}\right)^{2p}\right]
\nonumber\\ 
&\times \left|\frac{{ d}\ln\sigma}{{ d}\ln M}\right|\,
\exp\left(-\frac{q\delta_{sc}^2(z)}{2\sigma(M)^2}\right) dM \;,
\end{align}
where $\r_0$ is the mean comoving cosmological mass density, 
while $p=0.3$ and $A=0.32218$. 
The Press-Schechter mass function is recovered for $q=1$, $p=0$ and $A=0.5$ 
\cite{Bond et al. 1991}. 

In what follows, $\sigma(M)$ and $\delta_{sc}(z)$ are
computed according to the formulas 
\cite{Kitayama&Suto, Peacock&Dodds 1994, Sugiyama 1995}
\begin{align}
\sigma \propto (1+2.208m^d-0.7668m^{2d}+0.7949m^{3d})^{-2/(9d)} \;,
\label{variance}
\end{align}
where $d=0.0873$, $m\equiv M(\Gamma h)^2/(10^{12}M_\odot)$ and 
\begin{align} 
\Gamma=\Omega_{0\rm m} h 
\exp\left[-\Omega_{0\rm b}\left(1+\sqrt{2h}/\Omega_{0\rm m}\right)\right] \;. 
\end{align}
The quantities related to the density contrast are 
\begin{align}
\label{deltasc}
\delta_{sc}(z)=\frac{\delta_{c}\,D_+(z=0)}{D_+(z)} \;,
\end{align}
\begin{align}
\delta_{c}\approx\frac{3\,(12\pi)^{2/3}}{20}
(1+0.0123\log_{10}\Omega_{\rm m}) \;.
\end{align}
The linear growth factor of density fluctuations, 
normalized to unity at present, may be approximated as 
\cite{Carroll et al. 1992} 
\begin{align}
\label{Dplus}
D_+(z)=\fr{5\Omega_{\rm m}}{2(1+z)}
\left[\Omega_{\rm m}^{4/7}-\Omega_\Lambda
+\left(1+\fr{\Omega_{\rm m}}{2}\right)
\left(1+\fr{\Omega_\Lambda}{70}\right)\right]^{-1}
\end{align}
where $\Omega_{\rm m}=\Omega_{0\rm m}(1+z)^3/E^2(z)$, 
$\Omega_\Lambda=\Omega_{0\Lambda}/E^2(z)$, and
\begin{align}
\label{ez}
E(z)=H(z)/H_0=\left[\Omega_{0\rm m}(1+z)^3+\Omega_{0\Lambda}\right]^{1/2} \;.
\end{align}

\section{The stochastic GW background}

In order to evaluate the GW output generated by a spatial distribution
of CDM halos we will exploit Eq.~(\ref{MF}) which provides a good fit
to N-body simulations of structure clustering in a variety of cosmological
models, at least over the redshift range 
$z=0-4$ \cite{Sheth et al. 2001, Sheth&Tormen 2002, Sheth&Tormen99,
Kauffmann et al. 1999, Jenkins et al. 1998}. 

For theoretical consistency, we have chosen to follow the same strategy
adopted by Ref.~\cite{Sheth et al. 2001} as described in the previous section. 
Therefore, in our numerical computation, 
we consider CDM structures over a mass range 
$M=5 \times 10^9M_\odot-5\times 10^{15}M_\odot$, 
which virialize at redshifts from 
$z=0$ to $4$. Each of these structures is approximated by a homogeneous
ellipsoidal perturbation with mass $M$, linear mass variance 
$\sigma^2(M)$ and critical linear density contrast $\de(M,z)=B(\sigma^2,z)$; 
in other words, every perturbation represents the most probable ellipsoid 
($p=0$ and $e=e_{mp}$) of mass $M$ which collapses at redshift $z$. 

Given the density contrast, the ellipticity and the prolateness, we then
calculate the eigenvalues 
of the external tidal tensor, using Eqs.(\ref{ellipticity})-(\ref{density}) 
and the relation $\lambda^\pr_\a=\lambda_\a-\de/3$. 
Next, we linearly rescale all quantities to the initial redshift $z_i=40$, 
at which the ellipsoidal evolution of the density perturbation starts,
following Eqs.~(\ref{axis})-(\ref{const2}). 
In fact, while the mass function provides 
the number of halos virializing at a given redshift (in our case $z=0-4$), 
the evolution of matter density perturbations, giving rise to these virialized
objects, begins much before, {\it i.e.} at very high redshifts (in our case
$z_i=40$). 
The initial conditions on the scale factor are
given by the relation $a(z_i)=1/(1+z_i)$ 
and by the well-known Friedmann equations, while, as we have already
anticipated, the initial conditions on the axis lengths  
and their time derivatives are specified by the Zel'dovich approximation setup
\begin{align}
\label{Zeld1}
R_\a(z_i)=a(z_i)(1-\lambda_\a)
\end{align}
and
\begin{align}
\label{Zeld2}
\dot{R}_\a(z_i)=H(z_i)\left[R_\a(z_i)-a(z_i)f(z_i)\lambda_\a\right]\;,
\end{align}
where $f(z)\approx\Omega_{\rm m}^{0.6}+\left(1/70\right)
\left[1-1/2\Omega_{\rm m}\left(1+\Omega_{\rm m}\right)\right]$ is the
growth rate of density fluctuations ({\it e.g.} Ref.~\cite{Lahav}).

For each $M$ and $z$, using Eq.~(\ref{hab_quadrupole}) and switching to the
proper time $t$, we evaluate the two independent components of the
gravitational radiation produced by a CDM halo, assuming that it is casually
oriented and placed at
a comoving distance $r(z)$ from the observer, where $z$ is the collapse
redshift. In this way, we observe today the radiation emitted at the
virialization time when, according to our ellipsoidal model, the GW output has
the maximum value. 
Actually, adopting this strategy, we slightly 
underestimate the total GW background, 
since we do not take into account those CDM halos which are 
still away from virialization.    
Moreover it is noteworthy here that, in our approach, we
have extrapolated Eq.~(\ref{hab_quadrupole}) outside its range of validity. 
In fact, this formula holds on scales well inside the Hubble horizon and in 
the wave zone ({\it i.e.} at distances larger than both the characteristic 
wavelengths and the characteristic size of the source), while, as we previously
noticed, CDM halos generate gravitational waves whose frequency is comparable
with the inverse of the Hubble time.
Nonetheless, as shown in the next Section, our results agree with several 
analytic approximations and previous works.

To account for all the directions of observation, 
we convert the two independent
states $h^\a_{e{}\,\b}(t)$ of tensor polarization 
from the frame 
associated with the ellipsoid principal axes
to the observer frame, 
assuming that CDM structures emit in all directions and
are uniformly distributed all around the observer. For this purpose, we use
the relation $h^\a{}_\b(t,\Omega)=R^{T\,\a}{}_\nu(\Omega) \, 
h^\nu_{e{}\,\mu}(t)\, R^\mu{}_\b(\Omega)$ 
where $R^\a{}_\b(\Omega)$ is the general form of the 
rotation matrix with $\psi=0$ \cite{Goldstein} 
\begin{displaymath}
\label{rotation}
R^\a{}_\b(\Omega)\equiv
\left(\begin{array}{ccc}
\cos\phi & \sin\phi & 0\cr
-\cos\theta\sin\phi & \cos\theta\cos\phi & \sin\theta\cr
\sin\theta\sin\phi  & -\sin\theta\cos\phi & \cos\theta 
\end{array}\right)\;,
\end{displaymath}
and the solid angle $\Omega\equiv\left(\theta,\phi\right)$ 
is defined following the conventions of Ref.~\cite{Forward}. 

Since our aim is to estimate the energy density
\begin{align}
\label{Omega_gw}
\Omega_{\rm GW}(\n)\approx \fr{2\pi^2}{3H_0^2}\n^3\textrm{PSD} (\n)
\end{align}
associated with the stochastic GW background at the observer 
({\it e.g.} Ref.~\cite{Maggiore}), 
we need to know the power spectral density PSD$(\n)$, which one can obtain from
the Parseval's theorem as 
\begin{align}
\label{PSD}
\langle h^\a{}_\nu(t) h^\nu{}_\b(t)\rangle =
\int^\infty_{-\infty}d\n\textrm{PSD} (\n)\;. 
\end{align}
That depends on the redshifted proper frequency $\n=\n_e/(1+z)$, 
where $\n_e$ is the proper frequency at the emission time. In Eq.~(\ref{PSD})
angle brackets denote time averaging at a given spatial point.

Thus, we first numerically evaluate the PSD$(\n,z,M,\Omega)$ of each individual
component of $h^\a{}_\b(t,\Omega)$ at each fixed value of $M$, $z$ and
$\Omega$, then we average the calculated PSDs over all directions by
integrating over the solid angle and dividing by $4\pi$ and, finally, we sum
over the components in order to get a mean power spectral density PSD$(\n,z,M)$
for every $z$ and $M$.

Since in our model each CDM halo is approximated by a most probable ellipsoid
of mass $M$ which collapses at redshift $z$, we multiply each PSD$(\n,z,M)$ by
the number $dN(z,M)=n(z,M)dMdV$ of halos in the comoving volume 
$dV(z)=4c\,\pi r^2(z)dz/\left(a_0H(z)\right)$ where $c$ is the speed
of light and $a_0\equiv 1$ is the present value of the scale factor.
Finally, we insert the resulting quantity in the definition of the GW energy
density Eq.~(\ref{Omega_gw}) and integrate over all
redshifts and masses to obtain the total $\Omega_{\rm GW} (\n)$. 

All the results are presented in the next section. 

\section{Results}
\label{res.}

Our result concerning the GW output of each CDM halo, an example of 
which is given in Fig.~\ref{gws}, is consistent  
with previous works in this field \cite{Quilis98,Quilis00}.
Moreover it is comparable to analytic approximations 
({\it e.g.} \cite{grav,Shapiro}) as
\begin{align}
\label{analytic approx}
h\approx \fr{3 \times 10^{-11}}{D/100\rm Mpc}\times
\fr{GM \left(10^{15}\right)^{2/7}}{c^2L}\times  \fr{M}{10^{15}M_\odot},
\end{align}
where $h$ represents the amplitude of a GW signal coming 
from a non-spherically symmetric collapsing object with characteristic 
size $L$ at distance $D$ from the observer.

It is worth noting that the produced gravitational radiation has a very long
characteristic period, approximately given by the inverse of the halo evolution
time, which, according to the ellipsoidal model, 
corresponds to frequencies of the order of
$\n\approx10^{-18}$Hz. This excludes, therefore, any direct
detection of a complete pulse, but still allows for 
the possibility of GW detection via secondary CMB anisotropy and polarization 
and via the ``secular effect'' discussed in Refs.~\cite{Quilis98,Quilis00}.
The latter takes place
when a gravitational-wave crosses two testing particles; this induces
a variation in their relative distance which increases in time, 
since this effect lasts for many years.

Actually, besides what stressed in the previous Section, there are other 
reasons for which the ellipsoidal collapse approximation to CDM halo 
virialization underestimates amplitude and frequency 
characterizing the GW background. In fact, using this approach, 
the evolution of cosmic structures is regarded as a continuous phenomenon
which neglects merging effects and any possible features of variability that,
according to Ref.~\cite{Quilis98}, should be characterized by a
dynamical frequency of the order of $\n\approx10^{-17}$Hz. 

\begin{figure}[tbp]
\includegraphics[width=0.50\textwidth]{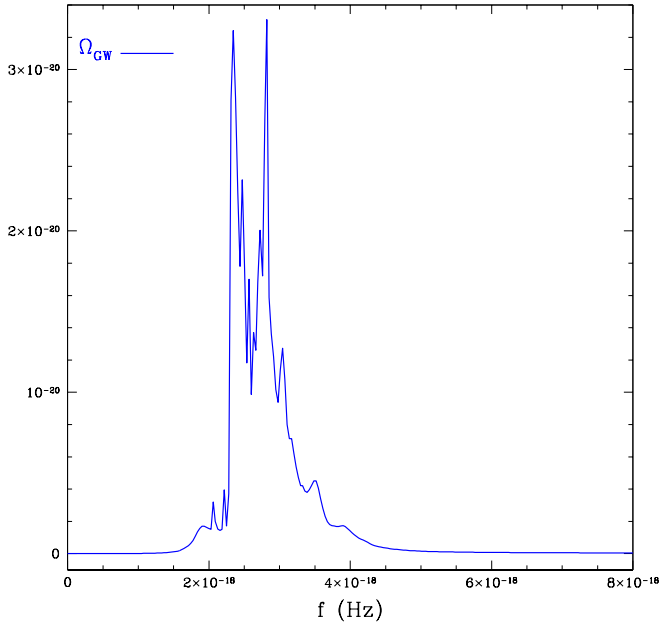}
\caption{The total energy density $\Omega_{\rm GW}(\n)$ associated with the
  stochastic GW background induced by CDM halos 
  as a function of the proper frequency $\n$ at observation.}
\label{Omega_gw_tot}
\end{figure}

In Fig.~\ref{Omega_gw_tot} the main result of our paper is shown, 
{\it i.e.} the
total energy density $\Omega_{\rm GW}(\n)\approx 10^{-20}$, 
associated with the stochastic halo-induced GW background, 
as a function of the proper 
frequency $\n$ at the observation. 
The total spectrum of the signal is composed by many 
single peaks which represent the contribution to the total background 
from each most probable halo weighted via the mass function
at different redshifts. On the other hand, as the following discussion shows,
these peaks are caused by the subset of structures leading to a non-negligible
GW signal.
In fact Eq.~(\ref{hab_quadrupole}) shows that the GW amplitude is
proportional to the inverse of the comoving distance, while, from the
expression of the efficiency $\epsilon=GM/(c^2 L)$ and the total radiated
energy $E_{GW}=\epsilon Mc^2$, where $L$ represents the characteristic halo
size at virialization, it follows that more massive objects give rise to
higher values of the GW strain. This effect is also confirmed by numerical 
estimates of the power spectral density for different objects in
the redshift range $0\le z\le 4$. In fact, masses of the order of $10^8-10^9
M_\odot$, although weighted via the mass function 
in Eq.~(\ref{MF}), contribute to 
$\Omega_{\rm GW}(\n)$ by only a factor of orders of $10^{-30}-10^{-28}$; 
since the amplitude of the gravitational waves decreases with distance, the 
greater is the redshift $z$, the lower is their contribution. 
Thus only a few peaks are visible in Fig.~\ref{Omega_gw_tot} since 
the energy density produced by less
massive structures is completely negligible with respect to the effect 
(of orders of $10^{-21}-10^{-20}$)
of far more massive objects ($10^{14}-10^{15} M_\odot$) at
low redshifts, $z\le 1$. 
\begin{figure}[tbp] 
\includegraphics[width=0.48\textwidth]{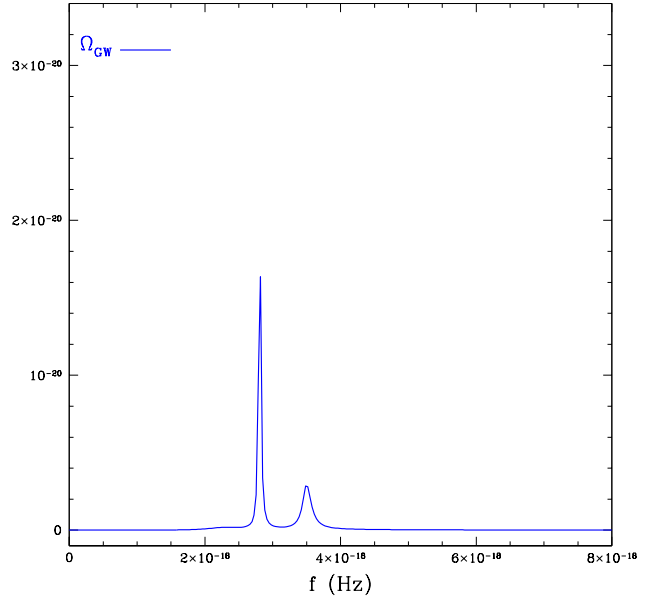}
\caption{Contribution to the total energy density $\Omega_{\rm GW}(\n)$ by two
  most probable CDM halos of mass $5\times 10^{15}M_\odot$ placed at
  redshifts $z=0.025$ and $0.075$.}
\label{mostcontribute}
\end{figure}
Consequently, the dominant contribution to the stochastic GW background 
is likely to be produced by CDM halos corresponding to 
nearby galaxies and galaxy clusters which 
contribute by several orders of magnitudes more than  
their substructures, although the latter are far more numerous.  Indeed, 
in  Fig.~\ref{mostcontribute} we may look at the 
contribution to the total $\Omega_{\rm GW}(\n)$ (see Fig.~\ref{Omega_gw_tot})
by two most probable halos of mass $5\times 10^{15}M_\odot$, placed at 
redshifts $z=0.025$ and $0.075$. In its maximum height, 
the signal reaches about half of the corresponding value 
in Fig.~\ref{Omega_gw_tot}. The remaining part of the signal is 
caused by many halos of comparable mass, as well as by those 
about one order of magnitude lighter, for which the mass decrease 
is compensated by the increase in the number. 
It is worth noting that the ellipsoidal collapse model introduces 
a three-peak pattern due to the freezing out method used to stop the axis 
collapse, which as a zero-th order approximation imposes stability at 
virialization, ignoring any residual dynamics. In the case of the 
specific geometrical configuration of the most probable ellipsoid 
considered in Fig.~\ref{mostcontribute}, this translates in two prominent 
peaks and a third negligible one. Actually, the residual dynamics at 
virialization would most probably imply a broadening of the spikes, 
decreasing the frequency splitting, possibly converging to a single 
peak for configurations close to sphericity. 

Finally and most importantly, the quantity $h^2\Omega_{\rm
GW}(\n)\approx 10^{-20}$ 
is comparable to the energy density associated with the
stochastic background induced by primordial GWs. 
In fact, if the energy scale of inflation is
$V^{1/4}\approx 1-2 \times 10^{15}\rm GeV$, 
the energy density associated with the 
primordial stochastic GW background, with a tensor spectral index
$n_T\approx0$, is $\approx10^{-21}-10^{-17}$ for frequencies of 
the order of $10^{-18}-10^{-17}$ Hz ({\it e.g} Ref.~\cite{Maggiore}). 

\section{Concluding remarks}

In this work, we have estimated the GW background from cosmological tensor
modes produced by the highly non-linear collapse of CDM density
perturbations, {\it i.e.} generated 
during the strongly non-linear stage of CDM halo
evolution. 

We found that the signal is significant at very low frequencies, 
$\n\approx 10^{-18}$ Hz, as a consequence of the cosmological 
time scales involved in the collapse of CDM halos. This signal appears 
as a broad peak made by the superposition of many impulses, 
all centered around frequencies of the order of $10^{-18}$ Hz. 
Most importantly, our results suggest that 
the signal is likely to be comparable to the primordial tensor power if 
inflation occurred at the GUT scale.
 
We want to stress that the homogeneous ellipsoidal collapse model, adopted
to simulate CDM halo evolution and virialization, underestimates the 
frequency and amplitude of the emitted gravitational waves, since, at each
redshift $z$, 
it does not take into account non-virialized objects and  
neglects variability features and merging effects that could enhance the 
anisotropic stress sourcing tensor modes, which are more sensible to the 
velocity field rather than to the peculiar gravitational potential. 
Consequently, the total energy density $\Omega_{\rm GW}(\n)$ generated 
by cosmic structures could even be of one or two order of magnitudes greater 
and overcome the stochastic background associated 
with primordial gravitational waves at the same frequencies (see also results
in Ref.~\cite{Quilis00}). 

The CDM halo GW background could also produce a non-negligible
contribution when considering the cosmological tensor-to-scalar ratio.

Due to the cosmological scales involved, and to the amplitude of the signal, 
it is reasonable to expect that these gravitational waves could affect the 
primary CMB anisotropies, contributing to the Integrated Sachs Wolfe (ISW)
effect caused by the time evolution of cosmological perturbations between us
and the last scattering surface. The stochastic GW background from CDM halos
might boost the temperature anisotropies on large angular scales, where 
however the contribution from density fluctuations dominates. 
On the other hand, the produced temperature quadrupole can be scattered 
off by the free electrons of the intra-cluster and intra-galactic media, 
giving rise to secondary E and B polarization modes similarly to what happens
for the primordial temperature quadrupole as described in
Ref.~\cite{Aghanim}. 
These contributions have 
to be taken into account when performing a precise evaluation of the level 
of CMB polarization anisotropy expected for the forthcoming polarization 
oriented CMB probes, in particular for what concerns the B modes. 

Most of these issues deserve a careful investigation in future works. 
Here we conclude stressing again our main results, suggesting that 
the amplitude of the stochastic GW background generated by CDM halos 
in their non-linear evolutionary phase is comparable or larger than 
the signal expected from the early universe in the inflationary scenario. 
We also remark that our findings are consistent with existing analytical
approximations. The forthcoming steps are the improvement of the 
calculation of the source of the signal, making use of cosmological N-body
simulations, as well as the computation of the induced CMB anisotropy in total
intensity and polarization. 

\acknowledgments
{We warmly thank Bepi Tormen for helpful discussions and suggestions.}

\appendix

\end{document}